\documentclass[conference]{IEEEtran}

\ifCLASSINFOpdf
  \usepackage[pdftex]{graphicx}
\else
   \usepackage[dvips]{graphicx}
\fi

\usepackage{amsmath}
\usepackage{amsfonts}
\usepackage{amsmath,epsfig,subfigure}
\usepackage{latexsym,amssymb}
\usepackage{cite,url}

\def\draftmode{}
\ifx\draftmode\undefined
\newcommand{\comment}[1]{}
\else
\newcommand{\comment}[1]{ \marginpar{$\Longleftarrow$}{\bf $<$#1$>$} }
\fi

\begin{document}
%
\title{A Hidden Markov Model for Localization Using Low-End GSM Cell Phones}

\author{\IEEEauthorblockN{Mohamed Ibrahim}
\IEEEauthorblockA{Wireless Intelligent Networks Center (WINC)\\
Nile University\\
Smart Village, Egypt\\
Email: m.ibrahim@nileu.edu.eg}\and \IEEEauthorblockN{Moustafa
Youssef}
\IEEEauthorblockA{Dept. of Comp. Sc. and Eng.\\
School of Elect. and Comp. Eng.\\
EJUST, Egypt\\
Email: moustafa.youssef@ejust.edu.eg} }


%


\maketitle

\begin{abstract}
Research in location determination for GSM phones has gained
interest recently as it enables a wide set of location based
services. RSSI-based techniques have been the preferred method for
GSM localization \emph{\textbf{on the handset}} as RSSI information
is available in all cell phones. Although the GSM standard allows
for a cell phone to receive signal strength information from up to
\textbf{seven} cell towers, many of today's cell phones are low-end
phones, with limited API support, that gives only information about
\textbf{the associated cell tower}. In addition, in many places in
the world, the density of cell towers is very small and therefore,
the available cell tower information for localization is very
limited. This raises the challenge of accurately determining the
cell phone location with very limited information, mainly the RSSI
of the associated cell tower. In this paper we propose a Hidden
Markov Model based solution that leverages the signal strength
history from \emph{\textbf{only the associated cell tower}} to
achieve accurate GSM localization. We discuss the challenges of
implementing our system and present the details of our system and
how it addresses the challenges. To evaluate our proposed system, we
implemented it on Android-based phones. Results for two different
testbeds, representing urban and rural environments, show that our
system provides at least 156\% enhancement in  median error in rural
areas and at least 68\% enhancement in median error in urban areas
compared to current RSSI-based GSM localization systems.

\end{abstract}



%
\IEEEpeerreviewmaketitle

\section{Introduction}
As cell phones become more ubiquitous in our daily lives, the need
for context-aware applications increases. Knowing the location of a
cell phone enables a wide set of location-based services including
navigation, location-aware social networking, and security
applications. Many technologies have been proposed to address the
localization problem in cell phones including the GPS system,
cellular-based system, and city-wide WiFi-based systems. GPS is
considered one of the most well known localization techniques
\cite{GPS99}. However, GPS is not available in many cell phones,
requires direct line of sight to the satellites, and consumes a lot
of energy. Therefore, research for other techniques for obtaining
cell phones' location has gained momentum fueled by both the user
needs for location-aware applications and government requirements,
e.g. FCC \cite{CELLUAR98}. City-wide WiFi-based localization for
cellular phones has been investigated in \cite{WiFi-City,PlaceLab2}
and commercial products are currently available \cite{SkyHook}.
However, WiFi chips, similar to GPS, are not available in many cell
phones and not all cities in the world contain sufficient WiFi
coverage to obtain ubiquitous localization. Similarly, using
augmented sensors in the cell phones, e.g. accelerometers and
compasses, for localization have been proposed in
\cite{CompAcc,AAMPL,Surroundsense}. However, these sensors are still
not widely used in many phones. On the other hand, GSM-based
localization, by definition, is available on all GSM-based cell
phones, which presents 80-85\% of today's cell
phones\cite{WIKIPEDIA_GSM}, works all over the world, and consumes
minimal energy in addition to the standard cell phone operation.
Many research work have addressed the problem of GSM localization
\cite{CELLUAR98,PlaceLab2,PlaceLab,GSM_INDOOR,CellSense}, including
time-based systems, angle-of-arrival based systems, and received
signal strength indicator (RSSI) based systems. Only recently, with
the advances in cell phones, GSM-based localization systems have
been implemented \cite{PlaceLab2,PlaceLab,GSM_INDOOR,CellSense}.
These systems are mainly RSSI-based as RSSI information is easily
available to the user applications.

According to the GSM standards, each cell phone can receive signals
from at most seven cell towers, one of them is the current tower the
cell phone is associated with and the others are the neighboring
cell towers. Current localization techniques for GSM networks are
designed to work with all these information. However, many cell
phones, even advanced ones, do not provide the API to get access to
the neighboring cell towers' information. This information is even
worse in many places in the world, especially in developing
countries, where low-end phones are the norm and cell towers'
density is very low. This brings up the challenge of providing
accurate GSM localization with minimal information, mainly the RSSI
form only the associated cell towers. Therefore, new techniques are
needed to address this new problem.

In this paper we propose a Hidden Markov Model (HMM)-based GSM
localization scheme using only the associated with cell tower
information. The main idea is to use the HMM to leverage the history
of the associated cell towers and their signal strength to obtain
accurate localization \footnote{Note that the associated cell tower
may change as the user moves.}. We describe the details of our HMM
and how we estimate its parameters and how given a sequence of RSSI
readings from the associated cell towers only we can estimate the
phone's location.

To evaluate our system, we implemented it on Android-enabled cell
phones, \emph{\textbf{using only the associated cell tower
information}}, and compare its performance to both deterministic
\cite{PlaceLab,GSM_INDOOR} and probabilistic \cite{CellSense}
fingerprinting techniques, and to Google's MyLocation service
\cite{MyLocation} under two different testbeds representing rural
and urban environments. Our results show that our system provides at
least 156\% enhancement in  median error in rural areas and at least
68\% enhancement in median error in urban areas compared to other
systems.

The rest of the paper is organized as follows:  Section
\ref{sec:background} gives a background on the current techniques
for RSSI-based localization in GSM networks. In Section \ref{HMM} we
discuss our new approach. Section \ref{results} presents the
performance evaluation of our system. Finally, Section
\ref{conclusion} concludes the paper and gives directions for future
work.

\section{Background}\label{sec:background}
This section presents a brief background on the current RSSI-based
techniques for GSM localization that we use for comparison with our
HMM-based technique, namely: cell-ID based techniques and
fingerprinting techniques.

\subsection{Cell-ID based Techniques}
Cell-ID based techniques, e.g. Google's MyLocation
\cite{MyLocation}, do not use RSSI explicitly, but rather estimate
the cell phone location as the location of the cell tower the phone
is currently associated with. This is usually the cell tower with
the strongest RSSI. Such techniques require a database of cell
towers' locations and provide an efficient, though coarse grained
localization method.

\subsection{Fingerprinting Techniques}
Fingerprinting based techniques store the RSSI signature of cell
towers at different locations in the area of interest in a database
during an offline phase. This database is searched during the
tracking phase for the closest location in the RSSI space to the
unknown location. Fingerprints are usually constructed by war
driving, where a car drives the area of interest continuously
scanning for cell towers and recording the cell tower ID, RSSI, and
GPS location for the associated and neighboring cell towers.

Current fingerprinting techniques for GSM localization are either
deterministic \cite{PlaceLab,GSM_INDOOR} or probabilistic
\cite{CellSense} techniques. Deterministic techniques does not take
the signal strength distribution into account. For example, each
location in the fingerprint of \cite{PlaceLab} stores a vector
representing the RSSI value from each cell tower heard at this
location. During the tracking phase, the K-Nearest Neighbors (KNN)
classification algorithm is used, where the RSSI vector at an
unknown location is compared to the vectors stored in the
fingerprint and the K-closest fingerprint locations, in terms of
Euclidian distance in RSSI space, to the unknown vector are averaged
as the estimated location. Probabilistic techniques, on the other
hand, store information about the RSSI distribution in the
fingerprint and try to estimate the most probable user location
during the online phase. For example, in CellSense \cite{CellSense},
the system stores the RSSI histogram for each cell tower at a
particular location and uses Bayesian-based inference to estimate
the user location.

Fingerprinting techniques require searching a larger database than
cell-ID based techniques but provide higher accuracy. Note that the
overhead of constructing the fingerprint is the same as constructing
the cell ID database as both require war driving.

\section{A HMM for GSM Localization}\label{HMM}
In this section, we present our HMM-based technique for GSM phones
localization using only the RSSI information from the associated
cell tower. We start by an overview of the system followed by the
details of the offline training and online tracking phases.

\subsection{Overview}
\begin{figure}[!t]
\centering
      \includegraphics[width=0.5\textwidth]{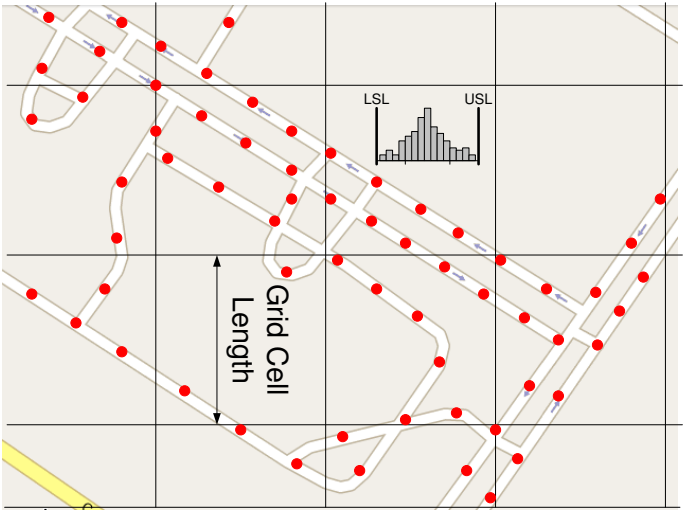}
  \caption{\textit{CellSense} approach for fingerprint construction. The area of interest is divided into grids and the histogram is constructed using the
  fingerprint locations inside the grid cell. No extra overhead is required for fingerprint construction. The grid cell length parameter can be used to tradeoff
  accuracy and scalability.}
  \label{fig:grid}
\end{figure}

We assume that the area of interest is divided into a grid as shown
in Figure \ref{fig:grid}. Our technique works in two phases: an
offline phase and and online tracking phase. The offline phase is
used to construct the HMM and estimate its parameters. As we
describe below in more details, each state represents a location in
the discrete physical space and an observation from a state
represents a RSSI reading from the associated cell tower. During the
online tracking phase, a sequence of observations, representing the
history of RSSI readings from the associated cell towers, is input
to the HMM to estimate the
 most probable sequence of states (locations). The last state in the most probable
sequence of states is used as the estimated location. In the following subsections, we give details about our system.

\subsection{Mathematical Model}
Without loss of generality, let $\mathbb{L}$ be a two dimensional
physical space where each location represent one grid cell. A HMM,
$\lambda$ can be represented as $\lambda= (S, V, A, B, \pi)$
\cite{}, where:

\begin{itemize}
\item $S=\{S_1,S_2,S_3,...,S_{N}\}$ is the set of possible states and $N= |S|$.
In our case, each state represents a grid location in the physical
space $\mathbb{L}$.

\item $V=\{v_1,v,2,v_3,...,v_M\}$ is the set of observations from the
model  and $M= |V|$. In our case, each observation is an ordered
pair of (Associate cell tower ID, RSSI).

\item $A= \{a_{ij}\}$ is the state transition probability
distribution, where $a_{ij}= P[q_{t+1}= S_j|q_{t}= S_i], i<i, j< N$
and $q_t$ is the state at time $t$.

\item $B= \{b_j(k)\}$ is the observation symbol probability
distribution in state $j$, where $b_j(k)= P[v_k\, \mathrm{at}\,
t|q_{t}= S_j], i<j< N, 1<k<M$ and $v_t$ is the output symbol at time
$t$.

\item $\pi=\{\pi_i\}$ is the initial state distribution, where $\pi_i=P[q_1=S_i]$.
\end{itemize}

 Therefore, the problem becomes, given a sequence of observations $O = (O_1,...,O_T)$, where $T$ is a system parameter and each $O_i \in V, 1<i<T$, we want to
find the most probable sequence of locations (states) $Q
=(q_1,...,q_T)$, where each $q_i \in S, 1<i<T$.

\begin{figure}[!t]
\centering
      \includegraphics[width=0.5\textwidth]{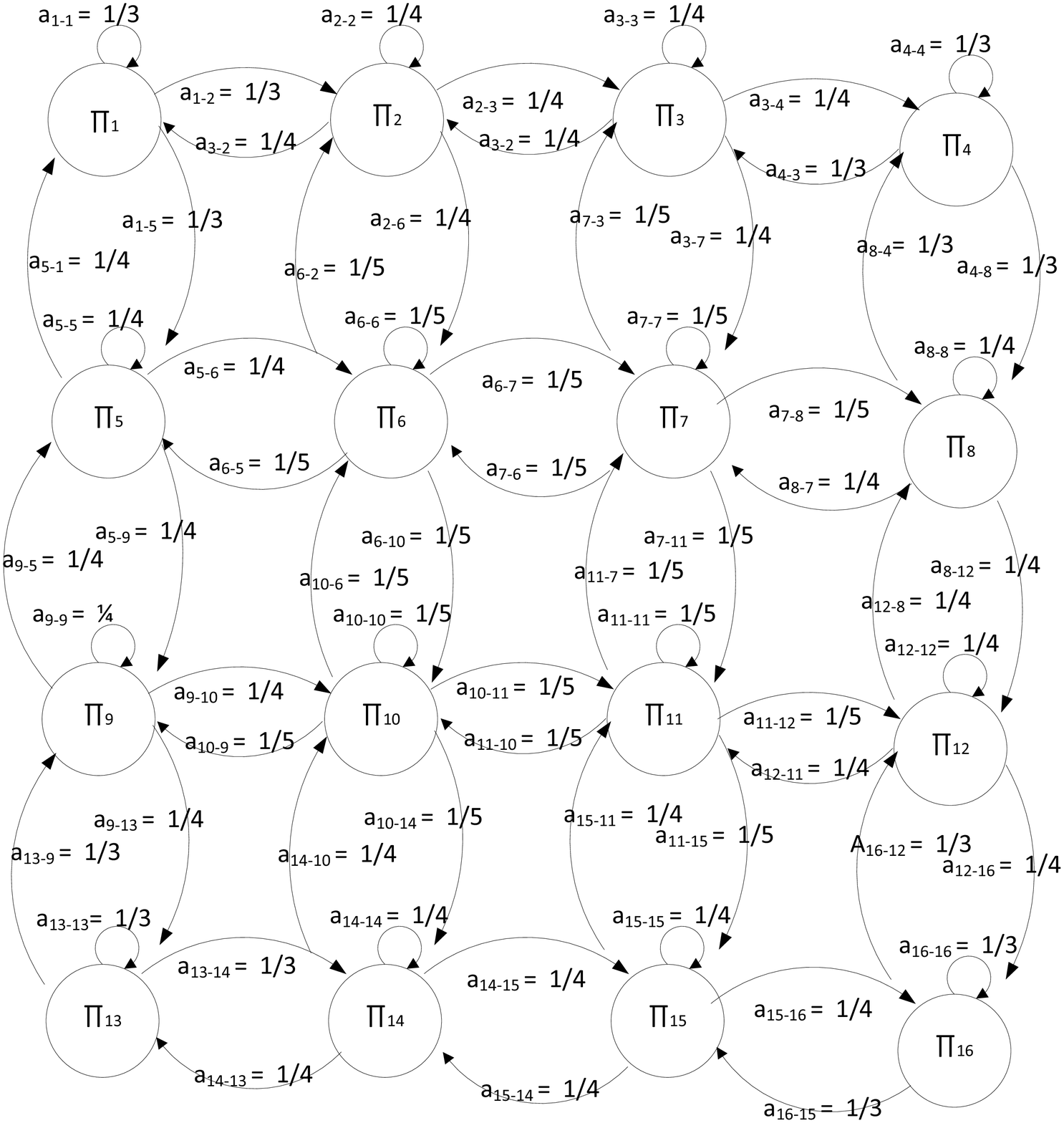}
  \caption{The equivalent HMM for Figure \ref{fig:grid}. There is a transition from each state (grid cell) to its four neighboring states only.}
  \label{fig:hmm}
\end{figure}

\subsection{Offline Phase}
The purpose of this phase is to construct the HMM and estimate its
parameters, namely $(S, V, A, B, \pi)$. For our GSM localization
system, the parameters are estimated as:
\begin{itemize}
  \item $S$: Since each state in our model represents a physical
  grid, the number of states, N, is the number of grid cells. Note
  that the grid cell size can be tuned to balance accuracy and
  complexity of implementation.

  \item $V$: At each state our set of observations corresponds to the different (Associated cell tower,
  RSSI) pairs that can be received inside this cell.

  \item $A$: To estimate the state transition matrix, we take
  advantage of the grid structure. Since a user can only move
  between adjacent grid cells, and each state represents a cell, the
  transition probability is taken to be unifrom over all the neighbors of a given cell and zero otherwise as shown in Figure
  \ref{fig:hmm}. Note that, in general, the transition matrix should
  depend on the shape of the road network. This can be obtained from
  digitized road maps, or for simplicity, it can be assumed that
  each cell is connected to its four or eight neighbors.

  \item $B$: To estimate the observation probability distribution at
  each cell, we use a technique similar to our previous work in
  \cite{CellSense}, where all the data points collected during war driving inside a given cell are used to estimate the RSSI histogram inside this cell.
  However, contrary to \cite{CellSense}, the histogram is constructed for the (Associated cell tower, RSSI)
  pairs and not for each individual cell tower, since in \cite{CellSense} we assume that we have information about all neighboring cell towers.

  \item $\pi$: If the initial state distribution is known, it can be used as is. If this information is not available, the steady state probability
  distribution, $\pi_{ss}$
  of the states can be used as an estimate for the initial state distribution. This can be estimated from the transition probability matrix, $A$
  as $\pi_{ss}A=\pi_{ss}$.
\end{itemize}

Once the HMM parameters are estimated, the system is ready for the
online tracking phase.

\subsection{Discussion}
The grid cell size parameter can be used to trade-off accuracy and
overhead. The larger the grid cell size, the lower the overhead as
the number of grid cells decreases. However, the grid cell size
increases reducing accuracy. We quantify the effect of the grid cell
size parameter on the system performance in the next section.

In order to estimate the observation probability distribution ($B$),
war driving is required. Since war driving is currently performed by
many entities, such as Google, there is no overhead for constructing
the observation probability distribution.
\subsection{Online Phase}
The idea of the technique is to use the history of RSSI values from
the associated cell tower to estimate the user location. In
particular, during the online phase, the user is moving in the area
of interest receiving signal strength information from the
associated cell tower only. Given a sequence of observations of
length $T$, $O = (O_1, ..., O_T)$, we want to find the location
where the user exists in at the end of the sequence. To estimate
this location, we compute the most probable sequence of states $Q=
(q_1, ..., q_T)$ given the sequence of observations seen by the user
using the Viterbi algorithm \cite{viterbi}. $q_T$ is returned as the
estimated user location. Note that increasing the observation
sequence length adds more information and hence should increase
accuracy. However, this comes with an increase in latency. We
quantify this tradeoff in the next section.


\section{Performance Evaluation}\label{results}
In this section, we study the effect of different parameters on our
system and compare its performance to other RSSI-based GSM
localization systems described in Section \ref{sec:background}.

\subsection{Data Collection}
We collected data for two different testbeds. The first testbed
covers the Smart Village in Cairo, Egypt which represents a typical
rural area. The second testbed covers a 5.5 Km$^2$ in Alexandria
representing a typical urban area. Data was collected using a
T-Mobile G1 phone which has a GPS receiver (used as ground truth for
location) and running the Android 1.6 operating system. Although the
phone provides information about all neighboring cell tower, we
\emph{\textbf{did not }}use this information in evaluating
the techniques to simulate the low-end cell phones.

We implemented the scanning program using the Android SDK. The
program records the (cell-ID, signal strength, GPS location,
timestamp) for the cell tower the mobile is connected to as well as
the other six neighboring cell towers information as dedicated by
the GSM specifications. The scanning rate was set to one per second.
Two independent data sets were collected for each testbed: one for
training and the other for testing. Table \ref{testbeds} summarizes
the two testbeds.

\begin{table}
\centering
\begin{tabular}{|p{0.42in}||p{0.45in}|p{0.55in}|p{0.55in}|p{0.6in}|}
\hline

Testbed & Area covered & Training set size & Testing set size& Avg. num. towers/loc.\\
\hline
 One (Rural)& 1.958Km$^2$ & 1198 & 301 &  5.16\\
\hline
Two (Urban)& 5.451Km$^2$ & 2890 & 1051 & 5.97\\
\hline


\end{tabular}
\caption{Comparison between the two testbeds.} \label{testbeds}
\end{table}

\subsection{Effect of Grid Cell Size}
\begin{figure}[!t]
    \centering
\includegraphics[width=0.50\textwidth]{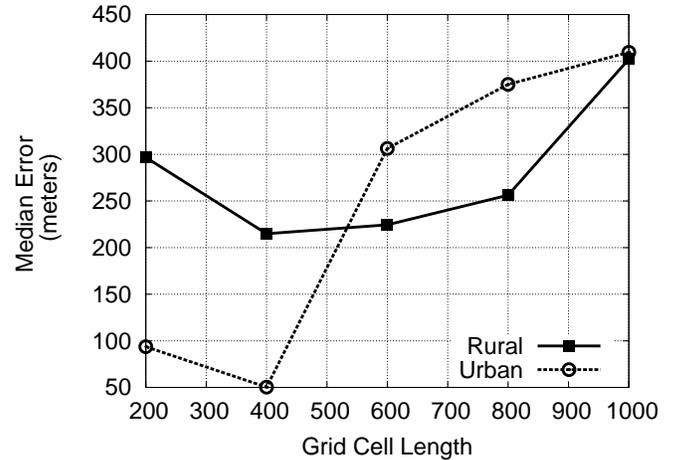}
            \caption{Effect of changing the grid cell length on our system's median error.}
    \label{fig:grid_cell}
\end{figure}

Figure \ref{fig:grid_cell} shows the effect of increasing the grid
cell length on accuracy. The observation sequence length parameter
was fixed at ten. The figure shows that as the grid cell length
increases, the median error decreases and then increases again. This
is attributed to two opposing factor: (a) As we increase the grid
cell size, we have a better estimate for the observation probability
distribution ($B$) as we have more samples inside each cell. This
has a positive effect on accuracy. (b) As we increase the grid cell
size, there is more error in location estimation inside each cell
which has a negative effect on accuracy.

On another hand, as we increase the grid cell size, the
computational overhead is reduced due to the decrease of the number
of cells. For the rest of the paper, we fix the grid cell size at
400 for the rural and urban cases as they provide the best accuracy.

\subsection{Effect of Observation Sequence Length}
\begin{figure}[!t]
    \centering
\includegraphics[width=0.50\textwidth]{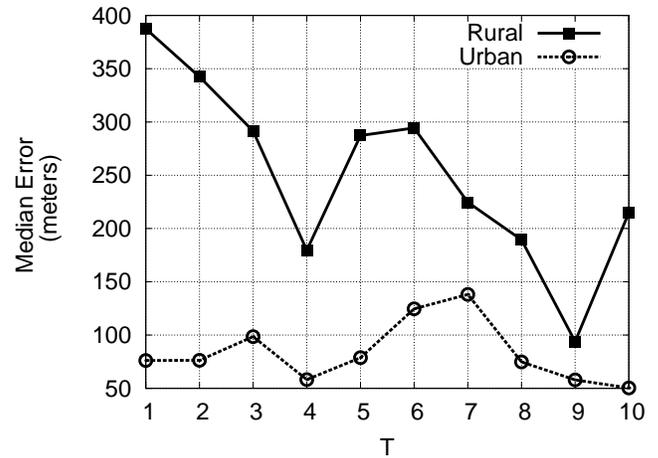}
            \caption{Effect of the observation sequence length ($T$) on our system's median error.}
    \label{fig:grid_W}
\end{figure}

Figure \ref{fig:grid_W} shows the effect of changing the length of
observation sequence, $T$, on the median error. The figure shows
that, as expected, as the window size increases, the accuracy
increases. This is due to increasing the amount of information that
can be used by the HMM model. However, increasing the length of the
observation sequence means that we need to wait for more samples
before estimating the user location, increasing the latency of
location estimation. Therefore, a balance has to be made between the
accuracy and latency depending on the required application.

%
%

\subsection{Comparison with Other Techniques}
\begin{figure}[!t]
\centering
    \subfigure[Testbed 1 (Rural). Grid Cell Length$= 400$, $T=9$.]{
      \includegraphics[width=0.5\textwidth]{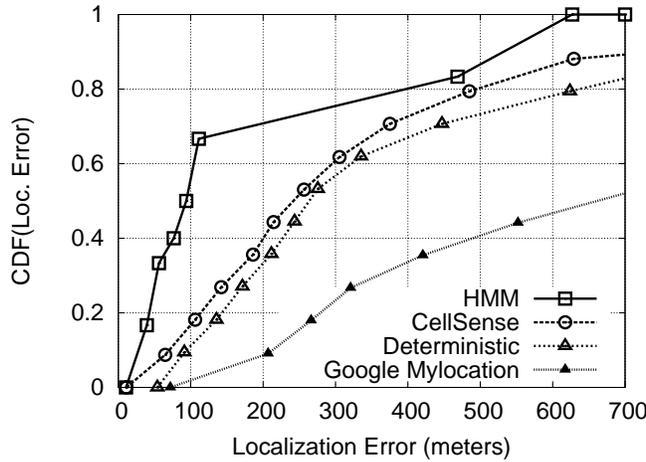}
    }
    \subfigure[Testbed 2 (Urban). Grid Cell Length$= 400$, $T=10$.]{
      \includegraphics[width=0.5\textwidth]{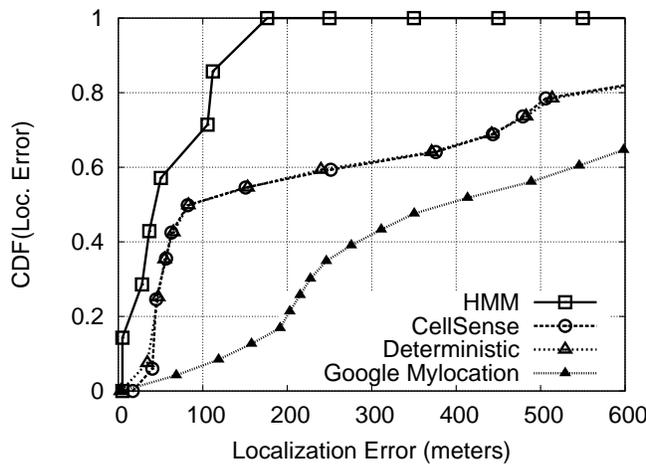}
    }
  \caption{CDF's of distance error for different techniques under the two testbeds. Note that the CDFs for techniques other than HMM has been truncated due to their heavy tail.}
  \label{fig:cdf}
\end{figure}

In this section, we compare the performance of our system, in terms
of localization error, to other RSSI-based GSM localization
techniques described in Section \ref{sec:background}. Figure
\ref{fig:cdf} shows the CDF of distance error for the different
algorithms for the two testbeds. The parameters that give the best
median error were used for \textbf{all} algorithms. Table
\ref{compare} summarizes the results. The table shows that our
system's accuracy is significantly better than any technique
achieving 93.85m median error in rural areas and 50.34m median error
in urban areas. This is better than any of the other techniques by
more than 156\% and 68\% for the rural and urban testbeds
respectively. All techniques perform better in urban areas than
rural areas due to the higher density of cell towers and the more
differentiation between fingerprint locations due to the dense urban
area structures. This is the same reason for the significant
enhancement in performance of our technique in rural areas.

\begin{table*}
\centering
\begin{tabular}{|p{2.0in}|p{1.1in}|p{0.9in}|p{0.9in}|p{0.5in}|}
\hline

Algorithms & Google's MyLocation& Deterministic & \textit{CellSense} & HMM\\
\hline Testbed 1-Rural Median Error(meters) & 656.37 (260.52\%)&
263.56 (599.38\%) & 240.76 (156.53\%)& 93.85\\
\hline
Testbed 2-Urban Median Error(meters)& 503.89 (900.9\%) & 89.12 (77.03\%)& 84.75 (68.36\%) & 50.34\\
\hline


\end{tabular}
\caption{Comparison between different techniques using the two
testbeds. Numbers between parenthesis represent percentage
degradation compared to the proposed HMM-based technique.}
\label{compare}
\end{table*}


\section{Conclusion}\label{conclusion}
We proposed a HMM-based localization system for GSM cell phones that
uses only the associated cell tower information. We presented the
details of the system and how it constructs the HMM model and
estimates its parameters and how we leverage the history information
to enhance the accuracy. We also implemented our system on
Android-based phones and compared it to other GSM-localization
systems under two different testbeds.  Our results show that our
system accuracy is better than all other techniques achieving 93.85m
median error in rural areas and 50.34m in urban areas, an
enhancement over the current GSM localization techniques of at least
156\% and 68\% for rural and urban areas respectively.




\section*{Acknowledgment}
This work is supported in part by a Google Research Award.




%



\bibliographystyle{IEEEtran} 
\bibliography{ref} 

\begin{thebibliography}{10}
\providecommand{\url}[1]{#1}
\csname url@samestyle\endcsname
\providecommand{\newblock}{\relax}
\providecommand{\bibinfo}[2]{#2}
\providecommand{\BIBentrySTDinterwordspacing}{\spaceskip=0pt\relax}
\providecommand{\BIBentryALTinterwordstretchfactor}{4}
\providecommand{\BIBentryALTinterwordspacing}{\spaceskip=\fontdimen2\font plus
\BIBentryALTinterwordstretchfactor\fontdimen3\font minus
  \fontdimen4\font\relax}
\providecommand{\BIBforeignlanguage}[2]{{%
\expandafter\ifx\csname l@#1\endcsname\relax
\typeout{** WARNING: IEEEtran.bst: No hyphenation pattern has been}%
\typeout{** loaded for the language `#1'. Using the pattern for}%
\typeout{** the default language instead.}%
\else
\language=\csname l@#1\endcsname
\fi
#2}}
\providecommand{\BIBdecl}{\relax}
\BIBdecl

\bibitem{GPS99}
P.~Enge and P.~Misra, ``{Special issue on GPS: The Global Positioning
  System},'' \emph{Proceedings of the IEEE}, pp. 3--172, January 1999.

\bibitem{CELLUAR98}
S.~Tekinay, ``{Special issue on Wireless Geolocation Systems and Services},''
  \emph{IEEE Communications Magazine}, April 1998.

\bibitem{WiFi-City}
Y.-C. Cheng, Y.~Chawathe, A.~LaMarca, and J.~Krumm, ``Accuracy characterization
  for metropolitan-scale wi-fi localization,'' in \emph{MobiSys '05:
  Proceedings of the 3rd international conference on Mobile systems,
  applications, and services}.\hskip 1em plus 0.5em minus 0.4em\relax New York,
  NY, USA: ACM, 2005, pp. 233--245.

\bibitem{PlaceLab2}
I.~Smith, J.~Tabert, A.~Lamarca, Y.~Chawathe, S.~Consolvo, J.~Hightower,
  J.~Scott, T.~Sohn, J.~Howard, J.~Hughes, F.~Potter, P.~Powledge,
  G.~Borriello, and B.~Schilit, ``Place lab: Device positioning using radio
  beacons in the wild,'' in \emph{Proceedings of the Third International
  Conference on Pervasive Computing}.\hskip 1em plus 0.5em minus 0.4em\relax
  Springer, 2005, pp. 116--133.

\bibitem{SkyHook}
{Skyhook wireless}, ``\url{http://www.skyhookwireless.com}.''

\bibitem{CompAcc}
R.~R.~C. Ionut~Constandache and I.~Rhee, ``Towards mobile phone localization
  without war-driving,'' in \emph{IEEE Infocom}, 2010.

\bibitem{AAMPL}
R.~S. Andrew~Offstad, Emmett~Nicholas and R.~R. Choudhury, ``Aampl:
  Accelerometer augmented mobile phone localization,'' in \emph{ACM MELT
  Workshop (with Mobicom 2008)}, 2008.

\bibitem{Surroundsense}
I.~C. Martin~Azizyan and R.~R. Choudhury, ``Surroundsense: Mobile phone
  localization via ambience fingerprinting,'' in \emph{ACM MobiCom}, 2009.

\bibitem{WIKIPEDIA_GSM}
\BIBentryALTinterwordspacing
{Wikipedia}, ``{Comparison of mobile phone standards --- {W}ikipedia{,} The
  Free Encyclopedia},'' [accessed 25-March-2010]. [Online]. Available:
  \url{http://en.wikipedia.org/wiki/Comparison_of_mobile_phone_standards}
\BIBentrySTDinterwordspacing

\bibitem{PlaceLab}
M.~Y. Chen, T.~Sohn, D.~Chmelev, D.~Haehnel, J.~Hightower, J.~Hughes,
  A.~Lamarca, F.~Potter, I.~Smith, and A.~Varshavsky, ``Practical
  metropolitan-scale positioning for {GSM} phones,'' in \emph{Proceedings of
  the Eighth International Conference on Ubiquitous Computing (UbiComp}.\hskip
  1em plus 0.5em minus 0.4em\relax Springer, 2006, pp. 225--242.

\bibitem{GSM_INDOOR}
A.~Varshavsky, M.~Y. Chen, E.~de~Lara, J.~Froehlich, D.~Haehnel, J.~Hightower,
  A.~LaMarca, F.~Potter, T.~Sohn, K.~Tang, and I.~Smith, ``Are {GSM} phones
  {THE} solution for localization?'' in \emph{WMCSA '06: Proceedings of the
  Seventh IEEE Workshop on Mobile Computing Systems \& Applications}.\hskip 1em
  plus 0.5em minus 0.4em\relax Washington, DC, USA: IEEE Computer Society,
  2006, pp. 20--28.

\bibitem{CellSense}
M.~Ibrahim and M.~Youssef, ``Cellsense: A probabilistic rssi-based gsm
  positioning system,'' in \emph{IEEE Globecom}, 2010.

\bibitem{MyLocation}
{Google Maps for Mobile}, ``\url{http://www.google.com/mobile/maps/}.''

\bibitem{viterbi}
G.~Forney~Jr, ``{The viterbi algorithm},'' vol.~61, no.~3, 1973, pp. 268--278.

\end{thebibliography}

\end{document}